\documentclass[twocolumn,aps,prl,amsfonts,amssymb,amsmath,showpacs,english]{
revtex4-1}
\usepackage[T1]{fontenc}
\usepackage[latin9]{inputenc}
\usepackage{amssymb}
\usepackage{esint}
\usepackage{babel}
\usepackage{graphicx}
\begin{document}


\title{Interaction-induced localization of fermionic mobile impurities in a Larkin-Ovchinnikov superfluid }

\author{Jian Li$^1$, Jin An$^{1,2}$ and C. S. Ting$^1$}

\affiliation{$^{1}$Texas Center for Superconductivity and Department of Physics, University of Houston, Houston,
Texas 77204, USA \\
$^{2}$National Laboratory of Solid State Microstructures and Department of Physics, Nanjing University, Nanjing 210093, China}

\begin{abstract}
We theoretically investigate the interplay between the fermionic mobile impurity atoms and a Larkin-Ovchinnikov (LO) superfluid in a two dimensional optical lattice. We find that the impurity atoms get localized and can form pairs when the interaction between the impurity atoms and the LO superfluid is strong enough. These features are due to the phenomena of self-localization whose underlying mechanism is revealed by an effective model. The impurity atoms with finite concentrations can drive the transition from a two-dimensional-checkerboard-like LO state to a quasi-one-dimensional-stripe-like one. Experimental preparations to observe these features are also discussed.
\end{abstract}

\pacs{71.55.-i, 05.30.Fk, 67.85.Lm}

\maketitle

\emph{Introduction}-Impurities or defects which are unavoidable in the real materials always affect the physical properties of these systems in different kinds of ways. The impurity problems are particularly important in the context of superconductors where investigations on the impurity effects can help us to understand the experimental anomalies, probe the pairing symmetries and reveal the competing electronic correlations\cite{Zhu}. The impurities or defects can also be introduced into the cold atomic systems either by establishing a fine-grained optical speckle field\cite{White} or trapping impurity atoms\cite{AS, Zip, Schmid}. Compared to the condensed matter system, these impurities or defects are easier to be controlled in a cold atomic system. The experimental achievement has attracted increasing theoretical interest on the impurity effects in a two-component superfluid Fermi gas\cite{Demler}. Despite the impurity effects in a superfluid Fermi gas share some common features with that in the superconductors, they could have their unique properties. Especially, considering the impurity atoms trapped in a Fermi gas are mobile, the effect induced by them is actually an interactive process which involves the interactions between the impurity atoms and the superfluid Fermi gas, making impressions on both of them.

Such an interactive process is particular interesting when the two component fermions have different populations. Specifically, by tuning the population imbalance in a superfluid Fermi gas\cite{Zwierlein, Partridge, Liao}, it has been found in experiments that the magnetization or population imbalance can coexist with the superfluid, indicating the possible existence of the Fulde-Ferrell-Larkin-Ovchinnikov\cite{FF, LO} state. Since it has been well established that the FFLO state is more sensitive to nonmagnetic impurities than conventional s-wave superconductors or superfluid\cite{imLO}, the mobile impurity atoms are more likely to induce exotic physical consequences in the FFLO state.

Here we focus on the interplay between the fermionic mobile impurity atoms
and a Larkin-Ovchinnikov(LO) superfluid in a two-dimensional optical lattice. We find that there exits a critical value of the interaction between the impurity and the LO superfluid, above which the extended impurity states become localized. The underlying mechanism of the localization can be captured by our proposed effective model. Paired states of impurity atoms can be formed by occupying the bonding and anti-bonding states simultaneously when the number of impurities is slightly larger than the saddle points of the LO superfluid. On the other hand, novel structures of the LO superfluid can be induced and a transition from a two-dimensional-checkerboard pattern of the order parameters to a quasi-one-dimensional-stripe one can be driven by finite concentration of impurity atoms. This provides an indirect method to modify the patten of the LO superfluid experimentally.

\emph{Model Hamiltonian}-We consider a two-component superfluid Fermi gas with
mobile impurities in a two dimensional optical lattice, which can be described by the following
Hamiltonian:
\begin{eqnarray}
&&H=H_{c}+H_{d}+H_{cd}\\
&&H_{c}=\sum_{ij,\sigma}(-t_{c,ij}-\mu_{\sigma}\delta_{ij})c_{i,\sigma}^{\dagger}c_{j,\sigma} -U_{c}\sum_{i}c_{i,\uparrow}^{\dagger}c_{i,\downarrow}^{\dagger}c_{i,\downarrow}c_{i,\uparrow},
\end{eqnarray}
 $H_{d}=-\sum_{<i,j>}t_{d,ij}d_{i}^{\dagger}d_{j}$ and $H_{cd}=U_{cd}\sum_{i}\hat{n}_{i}^{c}\hat{n}_{i}^{d}$. $H_c$ describes a superfluid Fermi gas, in which $c_{i,\sigma}^{\dagger}$ represents the creation
operator of the $c$-atom with spin-$\sigma$ at site $i$ and $U_c$ is the attractive interaction between $c$-atoms. $t_c$
is the hopping integral of $c$-atoms between two nearest neighboring sites.
$H_d$ describes the femionic mobile impurity atoms in which $t_{d}$ is the hopping integral between the nearest neighboring sites.
$H_{cd}$ is the interaction between the $c$-atoms and the impurities.
Here we consider the repulsive interaction with $U_{cd}>0$.

We decouple model (1) by introducing mean-field order parameters $\Delta_i=-U_c\langle c_{i,\downarrow}c_{i,\uparrow}\rangle$, $n^c_{i,\sigma}=\langle c^{\dagger}_{i,\sigma}c_{i,\sigma}\rangle$ and $n^d_{i}=\langle d^{\dagger}_{i}d_{i}\rangle$. Then we have equation for $c$ atoms:
\begin{eqnarray}
&\sum_{j}\left[\begin{array}{cc}
H^c_{ij\uparrow} & \Delta_{ij}\\
\Delta_{ij}^{\ast} & -H_{{ij\downarrow}}^c
\end{array}\right]\left[\begin{array}{c}
u_{j,n}\\
v_{j,n}
\end{array}\right]=E^c_{n}\left[\begin{array}{c}
u_{i,n}\\
v_{i,n}
\end{array}\right]&
\end{eqnarray}
and impurity atoms: $\sum_{j}H^d_{ij}w_{j,n}=E^d_{n}w_{i,n}$, where $H^c_{ij\sigma}=-t_{c,ij}+(U_{cd}n^d_{i}-\mu_{\sigma})\delta_{ij}$, $\Delta_{ij}=\Delta_i\delta_{ij}$
and $H^d_{ij}=-t_{d,ij}+U_{cd}n^c_{i}\delta_{ij}$.
The particle number per site and the on-site s-wave pairing order parameter are given by: $n^c_{i,\uparrow}=\sum_{n}u_{i,n}^{\ast}u_{i,n}\Theta(-E^c_{n})$,
$n^c_{i,\downarrow}=\sum_{n}v_{i,n}^{\ast}v_{i,n}\Theta(E^c_{n})$, $\Delta_{i}=-U_c\sum_{n}u_{i,n}v_{i,n}^{\ast}\Theta(-E^c_{n})$ and $n^d_{i}=\sum_{n=1}^{N_d}w_{i,n}^{\ast}w_{i,n}$
where $\Theta(x)$ is the step function and the magnetization is defined as $m_i=n^{c}_{i,\uparrow}-n^{c}_{i,\downarrow}$. $N_d=\sum_{i}n^{d}_{i}$ is the total number of the fermionic impurity atoms which occupy the lowest $N_d$ energy levels. In our calculation, we choose a lattice size of $32\times32$ with periodic boundary condition and use $t_c=1$ as the energy unit and lattice constant $a=1$ as the length unit.
\begin{figure}[t]
\includegraphics[width=3.2in] {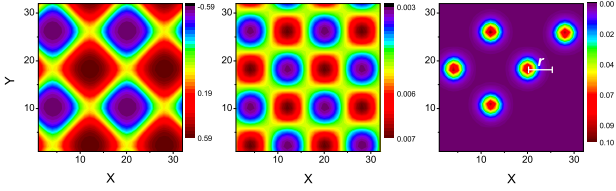}
\caption{(Color online) Contour-plot for (a): $\Delta_i$ without the impurities, (b) and (c): density distribution of five impurity atoms with $U_{cd}=1.0$ and $U_{cd}=4.0$, respectively. Here $t_d=0.3$ and $U_c=4.0$.}
\label{Fig1}
\end{figure}
The total particle number and population imbalance of the $c$-atoms is fixed to be $N^c_{total}=N^c_{\uparrow}+N^c_{\downarrow}=200$ and $N^c_{differ}=N^c_{\uparrow}-N^c_{\downarrow}=60$. Without impurities, as shown in Fig.1(a), we have an LO state for the $c$-atoms with a period of $\Delta_i$: $L=16$ along $x$ and $y$-direction, which is close to the estimation value $2\pi/q\approx17.8$ where $q$ is the difference between the Fermi surface radius of spin-up and spin-down $c$-atoms.

\emph{Interaction-induced localization}-Despite the LO state in Fig.1(a) breaks the translational symmetry of model (1), it preserves the $C_4$ symmetry of the optical lattice and has a new translational symmetry which can be characterized by $\overrightarrow{q}$ where $q_{x,y}=2\pi/16$. The density of the $c$-atoms $n_{i}^c$ has a two dimensional periodic distribution with local maximum (minimum) values corresponding to the peaks (saddle points) of $\Delta_i$.
For weak $U_{cd}$, the wave functions for the impurities are extended and the impurity density $n^{d}_{i}$ is periodically distributed, preserving the $C_4$ and "$\overrightarrow{q}$-translational" symmetry. As an example, we show $n^{d}_{i}$ for five impurity atoms in Fig.1(b).
However, once $U_{cd}$ exceeds a critical value $V_c$, the $C_4$ and "$\overrightarrow{q}$-translational" symmetry is broken and $n_{i}^{d}$ is limited in several small real-space ranges each of which has a radius $r$\cite{r} centered around the saddle points of the LO superfluid, as shown in Fig.1(c), suggesting that the impurity atoms are localized.
We find $V_c$ is weakly dependent of $N_d$ and the value of $V_c$ can be found from the plot of the inverse participation number $P^{-1}=\Sigma_{i}(n^{d}_{i})^2/(\Sigma_{i}n^{d}_{i})^2$ which characterizes the degree of the localization\cite{Wegner}.
As illustrated in Fig.2(a), for a single impurity atom, when $U_{cd}$ is smaller than $V_c$, $P^{-1}$ is almost zero while it gets significantly enhanced once $U_{cd}>V_{c}$. Accordingly, the radius $r$, labeled as blue circles in Fig.2(a), becomes smaller with the increasing of $P^{-1}$, characterizing a stronger localization. Notice the step-like data lines of $r$ in Fig.2 is due to the lattice effect. From the inset of Fig.2(a), we find that $V_c$ increases monotonically with $t_d$, indicating that a stronger interaction is needed to get a lighter impurity atom localized.
\begin{figure}[b]
\includegraphics[width=3.0in] {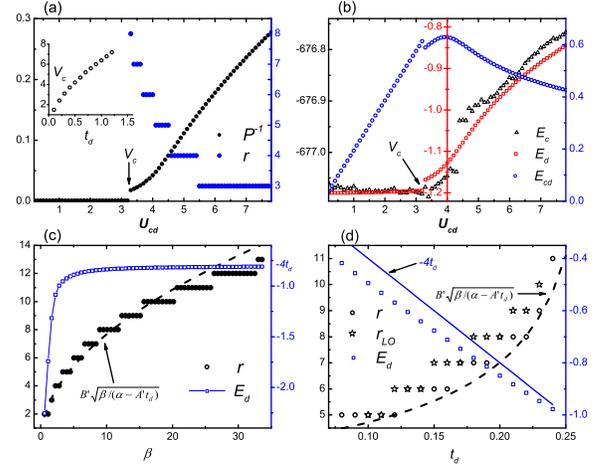}
\caption{(Color online) (a): Plot of $P^{-1}$ and the radius $r$ as the function of $U_{cd}$. Here $t_d=0.3$. Inset: Plot of $V_c$ as the function of $t_d$ for model (1).
(b): Plot of $E_c$, $E_d$ and $E_{cd}$ as the function of $U_{cd}$. (c) and (d): Plot of $r$ and total energy $E_d$ as the function of
(c): $\beta$ with $t_d=0.2$, $\alpha=-2.8$ and (d): $t_d$ with $\alpha=-3,\beta=10.8$ for model (4). In (d), $r_{LO}$ indicates the radius from model (1) with $U_{cd}=2.6$. In (c) and (d), the lattice size for model (4) is $50\times50$ and the fitting parameter are $A'=10.6$, $B'=2.0$.  Notice in all the figures the data lines correspond to the coordinates with the same color. }
\label{Fig2}
\end{figure}

To explore the origin of the localization, we plot all the three parts of the ground state energy as the function of $U_{cd}$ in Fig.2(b), in which $E_c$ is the energy of $c$-atoms, $E_d$ is the kinetic energy of the single impurity atom and $E_{cd}=\sum_{i}\langle U_{cd}n^c_in^d_i\rangle$ is the interaction energy between them. When $U_{cd}<V_c$, the energy of the $c$-atoms and the kinetic energy of the impurity atoms are almost unchanged while $E_{cd}$ increases linearly with the increasing of $U_{cd}$. Once $U_{cd}>V_c$ when the localization happens, $E_c$ and $E_d$ increases while $E_{cd}$ increases in a smaller rate then decreases with $U_{cd}$. This indicates that the extended state is favored by $E_c$ and $E_d$ while the localized state is favored by the interaction energy and the localization is a direct consequence of the competition between them.

Here the impurity atoms can get localized at
the positions where the potentials are distorted by themselves through the interaction $U_{cd}$. This can be interpreted as a process of self-localization\cite{RD, DKKD, KK}, which is in analogy with the formation of small polarons in condensed matter physics where the electrons can get self-localized by strong electron-phonon interaction\cite{polaron}. For an intuitive understanding of the self-localization of an impurity atom, we propose an effective model by assuming that the density of the $c$-atoms can be expanded as a function of $n^d_{i}$, i.e. $U_{cd}n^c_{i}=C_n-\alpha n^{d}_{i}+\beta (n^{d}_{i})^2+...$ where $C_n$ is a function of the density distribution of the $c$-atom in the absence of the impurity and does not change the nature of the self-localization qualitatively. By ignoring the irrelevant terms and up to the second order of $n^{d}_{i}$, we have the following effective model for the impurity atom:
\begin{eqnarray}
H_{eff}=-t_{d}\sum_{<i,j>}d_{i}^{\dagger}d_{j}-\sum_{i}[\alpha n^{d}_{i}-\beta (n^{d}_{i})^2]d_{i}^{\dagger}d_{i}.
\end{eqnarray}
where $\alpha,\beta>0$. The most important feature of model (4) is that the interaction part between the impurity and LO state in model (1) is reduced to an self-consistently determined external potential. This strategy is in analogy with the Kohn-Sham density functional theory\cite{KS}, although different physical origin is presented there. The main physics of the self-localization can be grasped by this effective model by reducing several competing energies to two parts: the first kinetic energy term and the second $\alpha$- or "external potential" term, where the former favors an extended state and the latter favors a localized one. There exists a critical value $\alpha_c$ at which the self-localization of the impurity atom occurs. A positive $\beta$ term stabilizes the localized state with finite localization length $\lambda$. To make an estimation on $\lambda$, we assume that the total energy of one impurity atom from model (4) should be $E_d=At_d/\lambda^{2}-B\alpha/\lambda^{2}+C\beta/\lambda^{4}$ where $A$, $B$ and $C$ are integral constant\cite{Gauss}. By minimizing the total energy we have the optimal localization length $\lambda=\sqrt{2C/B}\sqrt{\beta/(\alpha-A/Bt_d)}$ which is proportional to $r$\cite{r2}. From Fig.2(c)-(d) we can see that the numerical results for $r$ as the function of $\beta$ and $t_d$ are well fitted by our analytical estimation. With the increasing of $r$, the localization becomes weaker and the energy of the localized atom $E_d$ is closer to the extended case with energy $-4t_d$. We also plot the variance of the radius from model (1) (labeled as $r_{LO}$) as the function of $t_d$ in Fig.2(d) and find its behavior is qualitatively consistent with the results from our effective model.

\emph{Paired state}-When the impurity number is slightly larger than the saddle points of the LO state, they could form pairs.
In Fig.3(a), we plot the density distribution of nine impurity atoms. We can see that seven atoms are localized at different saddle points separately, while two atoms form a pair at one saddle point, which can be identified by their energy levels and wave functions. From Fig.3(b) we can see that
 \begin{figure}[b]
\label{Fig3}
\includegraphics[width=3.0in] {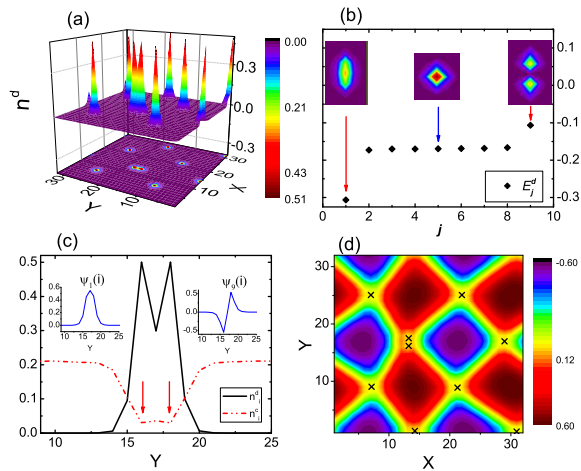}
\caption{(Color online) (a): 3D plot of $n_{i}^{d}$ for nine impurity atoms with $t_{d}=0.2$ and $U_{cd}=5.0$.
(b): The energy levels of the impurity atoms in (a).
Inset: $|\Psi_{1,5,9}(i)|^2$.
(c): 1D cut-plot of $n_{i}^{c}$ and $n_{i}^{d}$ for the paired states along the long axis. Inset: $\Psi_1(i)$ and $\Psi_9(i)$.
(d): $\Delta_i$ for the case in (a). Notice the positions of the impurity atoms are labeled as "$\times$".}
\end{figure}
the paired states occupy the highest and lowest energy level, and their probability densities $|\Psi_{1,9}(i)|^2$ are also different from the unpaired states, as shown in the inset-plots of Fig.3(b). Further calculations (not shown here) indicate that the paired state also forms on a lattice with bigger size, excluding the possibility of the finite size effect.

The pair formed by two impurity atoms is also the consequence of the interactive process between the impurities and LO superfluid. To see this, we make the 1D-cut-plot of $n^{c}_{i}$ and $n^{d}_{i}$ for the paired state along its long axis. From Fig.3(c) we can see that, different from two fermions trapped in a "hard" potential well, two impurity atoms here modify the density distribution of $c$-atoms and create a double-well structure (labeled as red arrows) through the interaction $U_{cd}$. The inter-well tunneling splits the energy into two levels which has the wave function of bonding and anti-bonding state, respectively. Due to their fermionic nature, two impurity atoms occupy the bonding and anti-bonding state simultaneously and their wave functions $\Psi_{1,9}$ can be well fitted with $(\phi_L\pm\phi_R)/\sqrt{2}$ where $\phi_{L,R}$ is the wave function of the individual well without tunneling, as exhibited in the inset of Fig.3(c).
Compared to the unpaired ones, the paired impurity atoms induce stronger local distortion to the LO superfluid, as shown in Fig.3(d), which is unfavored by the $c$-atoms. Therefore the impurity atoms do not always form paired states since they may not be supported by the total energy.

\emph{Impact of mobile impurity atoms on the LO superfluid}-
\begin{figure}[b]
\label{Fig4}
\includegraphics[width=3.2in] {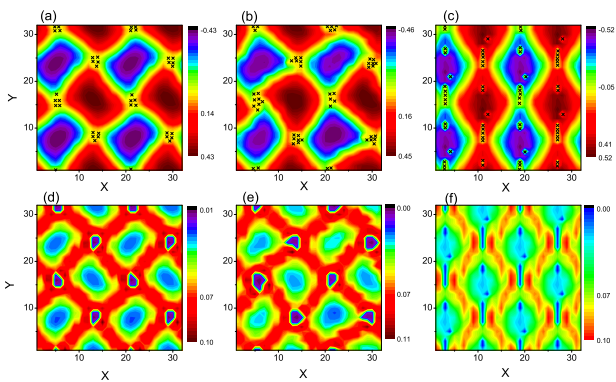}
\caption{(Color online) (a)-(c): Contour-plot of $\Delta_i$ of the LO state with (a): $N_{d}=40$, $U_{cd}=2.5$, (b) $N_{d}=51$, $U_{cd}=3.5$ and (c) $N_{d}=64$, $U_{cd}=5.0$.
(d)-(f): Contour-plot of $m_i$ with the same parameters with (a), (b) and (c), respectively.
Here $t_{d}=0.02$ and $U_c=3.7$. }
\end{figure}
As demonstrated in Fig.4(a)-(c) where $N_d=40,51,64$, the LO states are significantly distorted. Although it is not unusual to find the symmetry reduction of the LO state caused by the mobile impurities here, a surprising fact is that in all the three cases LO state maintain its original period to a certain degree or has a "quasi-$\overrightarrow{q}$-translational" symmetry. This phenomena identifies an important difference between the mobile impurities and general fixed-impurities, which are randomly distributed and always destroy the "$\overrightarrow{q}$-translational" symmetry of the LO state\cite{QianW}.
From Fig.4(b) we can see that the symmetry of the LO superfluid with $N_d=51$ is lower than the case of $N_d=40$. This is because the number of the impurity atoms $N_d=51$ does not support a commensurate structure of impurity atoms with the LO state, resulting in more distortions for the LO superfluid.
Compared to the former two cases, the pairing order parameters of the superfluid with $64$ impurity atoms show a quasi-one-dimensional-stripe-like pattern with non-crossing nodal lines. This indicates that the mobile impurity atoms can also affect the pattern of the LO superfluid besides the Zeeman field and temperature\cite{YMHS}.

Since both the polarized $c$-atoms and the impurity atoms are favored to stay at the saddle points, frustrations caused by their competition to the structure of the magnetization are inevitable, and thus novel distortions of the LO state can be expected for the cases of high impurity densities.  On the other hand, LO states with different patterns caused by the impurity atoms (e.g.,the 2D checkerboard and quasi-1D pattern) can be distinguished by detecting the spatial distribution of the population-imbalance experimentally.
As shown in Fig.4(d)-(e), for a 2D-like LO state, although suppressed by the impurity atoms, the local maximums of the magnetization still have crossing points which correspond to the saddle points of $\Delta_i$. For a quasi-1D LO state, the magnetization at the previous crossing points are totally destroyed and the polarized $c$-atoms are squeezed into non-crossing nodal lines, as shown in Fig.4(f).

When $U_{cd}$ is increased, the impurity atoms tend to localize near the saddle points or nodal lines, and repel out the local c-atoms of the background to minimize the interaction energy. This is similar to the picture of phase separation. A pure phase separation where impurities and superfluid occupy spatially different regions may be possible in cases with large impurity number and very large $U_{cd}$, which is beyond the scope of the present work. With and without impurities, the superfluid state is found to be LO state, whereas other states, like FF state and Sarma state, are never found stable in the parameter space we studied. Although there is no long range order in a pure 2D infinite system at finite temperatures, our results should be observable at finite temperatures because the cold atoms are confined in a 2D optical lattice and thus may form a quasi-2D system. In such a system the fluctuation effect is much suppressed and our mean field theory is expected to be valid.

\emph{Proposal for experimental setup}-Experimentally, Fermi-Fermi mixture was realized by using $^6$Li and $^{40}$K\cite{mixture} and the former was used to realize the possible FFLO state\cite{Zwierlein, Partridge, Liao}. Therefore two component $^6$Li atoms with spin-imbalance can be used as the background, i.e., the $c$-atoms, while one component $^{40}$K atoms can be treated as the impurities. Considering $^{40}$K is much heavier than $^6$Li, the critical value of $U_{cd}$ for localization should be relatively small and our mean-field treatment is applicable.

\begin{acknowledgements}\emph{Acknowledgements}-We thank Yan Chen, Matthias J. Graf and Xiaoling Cui for helpful discussions. This work was supported by the Texas Center for Superconductivity at the University of Houston and by the Robert
A. Welch Foundation under Grant No. E-1146. Jin An was also supported by NSFC(China) Project No.1117416.
\end{acknowledgements}

\emph{Note added}-After completion of this work, we noticed a similar work in which the velocity of a mobile impurity was investigated in one
dimension\cite{MADA}.


\begin{thebibliography}{99}

\bibitem{Zhu}A. V. Balastky, I. Vekhter, and Jian-Xin Zhu, Rev. Mod.
Phys. \textbf{78}, 373(2006);
H. Alloul, J. Bobroff, M. Gabay,and P. J. Hirschfeld, \emph{ibid}. \textbf{81}, 45 (2009).

\bibitem{White} M. White et al, Phys. Rev. Lett. \textbf{102}, 055301(2009).

\bibitem{AS} A. Schirotzek, C-H Wu, A. Sommer, and M W. Zwierlein , Phys. Rev. Lett. \textbf{102}, 230402(2009).

\bibitem{Zip} C. Zipkes et al, Nature(London) \textbf{464}, 388(2010).

\bibitem{Schmid}S. Schmid, A. Härter, and J. H. Denschlag, Phys. Rev. Lett. \textbf{105}, 133202 (2010).

\bibitem{Demler} E. Vernier, D. Pekker, M. W. Zwierlein, E. Demler, Phys Rev A \textbf{83}, 033619(2011);
        Y. Ohashi, \emph{ibid}. \textbf{83}, 063611(2011);
        L. Jiang, L. O. Baksmaty, H. Hu, Y. Chen, H. Pu, \emph{ibid}. \textbf{83}, 061604(R)(2011);
        M. Jiang et al, cond-matt/1109.1622.
        J. Li and C. S. Ting, Phys. Rev. B \textbf{85}, 094520 (2012)

\bibitem{Zwierlein}M. W. Zwierlein et al, Science \textbf{311}, 492 (2006).

\bibitem{Partridge}G. B. Partridge et al., Science \textbf{311}, 503 (2006);
 G. B. Partridgeet al., Phys. Rev. Lett. \textbf{97}. 190407 (2006).

\bibitem{Liao} Liao et al, Nature(London) \textbf{467}, 567(2010).

\bibitem{FF} P. Fulde and R. A. Ferrell, Phys. Rev. \textbf{135} A550(1964).

\bibitem{LO} A. I. Larkin and Y. N. Ovchinnikov, Zh. Eksp. Theor. Fiz, \textbf{47}, 1136(1964){[}Sov.
Phys. JETP \textbf{20}, 762(1965){]}.

\bibitem{imLO}L. G. Aslamazov, Sov. Phys. JETP \textbf{28}, 773 (1969);
              S. Takada, Prog. Theor. Phys. \textbf{43}, 27 (1970);
              L. N. Bulaevskii and A. A. Guseinov, Sov. J. Low Temp. Phys. \textbf{2}, 140 (1976);
              D. F. Agterberg and K. Yang, J. Phys. Condens. Matter \textbf{13}, 9259 (2001).
              Q. Cui and K. Yang, Phys. Rev. B \textbf{78}, 054501 (2008);
              Y. Yanase, New J. Phys. \textbf{11}, 055056 (2009).

\bibitem{Wegner}F. Wegner, Z. Phys. B\textbf{36}, 209(1980).

\bibitem{r} Throughout the paper, we define $r$ as the length from the center of the region occupied by the density of one impurity atom to
the site where $n^{d}_{i}<10^{-4}$.

\bibitem{RD}R. M. Kalas and D. Blume, Phys. Rev. A \textbf{73}, 043608 (2006);
F. M. Cucchietti and E. Timmermans, Phys. Rev. Lett. \textbf{96}, 210401 (2006);
K. Sacha and E. Timmermans, Phys. Rev. A \textbf{73}, 063604 (2006).

\bibitem{DKKD} D-S Lühmann, K. Bongs, K. Sengstock, and D. Pfannkuche, Phys. Rev. Lett. \textbf{101}, 050402 (2008).

\bibitem{KK}K. Targo\'{n}ska and K. Sacha  Phys. Rev. A \textbf{82}, 033601 (2010).

\bibitem{polaron}Charles Kittel, Introduction to Solid State Physics, 8rd. ed. (Wiley, New York, 2005), Chap. 14.

\bibitem{KS}W. Kohn and L. J. Sham, Phys. Rev. \textbf{140}, A1133(1965).

\bibitem{Gauss} As an example, for a variational localized wave function with Gaussian distribution $\phi(x,y)=e^{-\frac{x^2+y^2}{2\lambda^2}}$, we have $A=1$, $B=1/(4\pi)$ and $C=1/(9\pi^2)$.

\bibitem{r2} According to our defination of $r$, we have $r=\sqrt{ln10^{4}}\lambda\simeq1.61\sqrt{\beta/(\alpha-4\pi t_d)}$ for the Gaussian trial function.

\bibitem{QianW} Q. Wang, C-R Hu, and C. S. Ting, Phys. Rev. B \textbf{75}, 184515 (2007);

\bibitem{YMHS} Y. Matsuda and H. Shimahara, J. Phys. Soc. Jpn. \textbf{76}, 051005 (2007).

\bibitem{mixture}M. Taglieber et al, Phys. Rev. Lett. \textbf{100}, 010401 (2008); E. Wille et al, Phys. Rev. Lett. \textbf{100}, 053201 (2008);
T. G. Tiecke et al, Phys. Rev. Lett. \textbf{104}, 053202 (2010).

\bibitem{MADA}M. Schecter, A. Kamenev, D. M. Gangardt, and A. Lamacraft, Phys. Rev. Lett. \textbf{108}, 207001 (2012).

\end{thebibliography}
\end{document}